\documentclass[a4paper]{jpconf}
\usepackage{graphicx}
\begin{document}
\title{Dispersion of orbital excitations in 2D quantum antiferromagnets}

\author{Krzysztof Wohlfeld$^1$, Maria Daghofer$^1$, Giniyat Khaliullin$^2$, Jeroen van den Brink$^1$}

\address{$^1$IFW Dresden, P. O. Box 27 01 16, D-01171 Dresden, Germany}

\address{$^2$Max-Planck-Institut f\"ur Festk\"orperforschung,
              Heisenbergstrasse 1, D-70569 Stuttgart, Germany}

\ead{k.wohlfeld@ifw-dresden.de}

\begin{abstract}
We map the problem of the orbital excitation (orbiton) in
a 2D antiferromagnetic and ferroorbital ground state 
onto a problem of a hole in 2D antiferromagnet.
The orbiton turns out to be coupled
to magnons and can only be mobile on a strongly renormalized
scale by dressing with magnetic excitations. 
We show that this leads to a dispersion relation reflecting the
two-site unit cell of the antiferromagnetic background, in contrast to
the predictions based on a mean-field approximation and linear
orbital-wave theory.
\end{abstract}

\section{Introduction}
Calculating collective excitations in nonfrustrated but interacting spin 
systems may be considered as one of the few extremely successful stories in solving the strongly correlated electron systems.
Independently on whether the approach which is used to calculate
such excitations stems from the linear spin wave theory (for ordered
states) or the Bethe Ansatz solutions (for quantum disordered states
in 1D), a remarkable qualitative and quite often also quantitative agreement
with a number of experimental results is usually achieved, see e.g.~\cite{Col01}.

On the other hand, collective excitations in interacting {\it orbital}
systems have attracted much less attention of the community, primarily
due to experimental
problems in detecting the collective orbital excitations (orbitons)
\cite{Gru02}. {\it A priori}, excitations in the orbital degree of
freedom of Mott insulators should behave in a similar manner as those
of the spin sector:
in both cases the large Coulomb repulsion in the Mott insulators
`freezes' the charge degrees of freedom and largely reduces
the effective low energy Hilbert space so that instead of electrons it
is given by 
spin or orbital (pseudospin) degrees of freedom.
In place of the `spin-wave theory' for magnetic excitations, one
would then use `orbital-wave theory'~\cite{vdB98}.

It has recently been shown~\cite{Woh11}, however, that this
orbital-wave picture does not hold in many realistic systems. 
Here, we prove that in a generic 2D Kugel-Khomskii (KK) spin-orbital model \cite{KK1982}
with the antiferromagnetic (AF) and ferroorbital (FO) ground state the 
linear orbital wave theory breaks down and one needs to explicitly
take care of the coupling between spins and orbitals.
Furthermore, we show that for realistic parameters 
of the KK Hamiltonian, the dispersion of the coherent
excitation changes qualitatively and reflects the two-site unit cell
of the antiferromagnetic background, in contrast to expectations based
on linear orbital-wave theory. 

\section{Model and results}
The Hamiltonian of the generic 2D KK model that we want to study
describes a two-orbital system with orbital-conserving hopping and reads
\begin{equation}
\label{eq:h}
H\!=\! 2 J \sum_{\langle {\bf i}, {\bf j} \rangle }  \left({\bf S}_{\bf i }\! \cdot \!{\bf S}_{\bf j} \!+\!\frac{1}{4}\right) 
                               \left( {\bf T}_{\bf i } \!\cdot \!{\bf T}_{\bf j} \! +\! \frac{1}{4} \right) 
        \! +\! E_z \sum_{\bf i} T^z_{\bf i}, 
\end{equation}
where ${\bf S}$ (${\bf T}$) are the spin (pseudospin) operators which fulfill 
the SU(2) algebra and $J$ is the superexchange constant which describes 
the interaction of $S=1/2$ spins and $T=1/2$ (orbital) pseudospins.
Besides, $E_z$ is the crystal field which is implicitly assumed to be 
so large that in the ground state $|\phi \rangle$ of $H$ only one type
of orbital is occupied (the FO order of orbitals) and consequently
there is long-range AF order in temperature $T=0$.

\begin{figure}[t!]
\includegraphics[width=22pc]{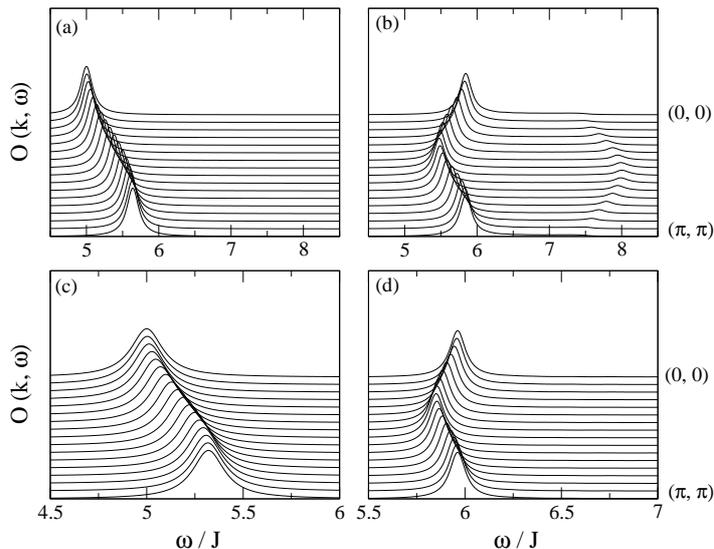}\hspace{2pc}%
\begin{minipage}[b]{14pc}\caption{\label{fig:1}Spectral function for
    the orbital excitation $O({\bf k}, \omega)$
    [Eq. (\ref{eq:spectral})] calculated in 
the linear orbital wave picture [(a) and (c)] and using the exact
mapping onto the effective $t$-$J$ model [(b) and (d)]. 
Upper (lower) panels for equivalent (inequivalent) values of the
hoppings between the two orbitals, see text. Note the different 
scales on upper and lower panels and that the incoherent spectrum in
panel (d) is extremely small and not visible at this
scale. Parameters: $E_z=5J$ 
and broadening $\eta = 0.05 J$.}
\end{minipage}
\end{figure}

The purpose of this study is to calculate the orbital excitations from the ground state  $|\phi \rangle$ of 
Hamiltonian (\ref{eq:h}), as defined by the spectral function $O ({\bf k}, \omega)$ :
\begin{equation} \label{eq:spectral}
O({\bf k}, \omega)=\frac{1}{\pi} \lim_{\eta \rightarrow 0} 
\Im \left\langle \phi \left| T^-_{\bf k} 
\frac{1}{\omega + E_{\phi}  - H -
i \eta } T^+_{\bf k} \right| \phi \right\rangle,
\end{equation}
where $E_{\phi}$ is the ground state energy and $ T^+_{\bf k}$ is the momentum-dependent orbital 
raising operator. 

In the `standard' orbital-wave approach \cite{Kha97,Ole05}, this spectral
function can be easily calculated (see, e.g., Ref. \cite{Woh11}) 
by mean-field decoupling of spins and orbitals and by: (i) introducing the Holstein-Primakoff
bosons for orbital pseudospins (cf. Ref. \cite{Woh09}) and (ii) neglecting interactions between
bosons. We then obtain:
\begin{equation}
\label{eq:spectralow}
O ({\bf k}, \omega)= \delta \left[ \omega - E_z + \frac{1}{2}zJ_{\rm OW}(1-\gamma_{\bf k}) \right], 
\end{equation}
with $z=4$ is the coordination number in 2D and $\gamma_{\bf k}=(\cos k_x + \cos k_y) / 2$ 
is the lattice structure factor in 2D. Here $J_{\rm OW} = 2J \langle \tilde{\phi} | 
{\bf S}_{\bf i } \cdot {\bf S}_{\bf j} + \frac{1}{4} | \tilde{\phi} \rangle \simeq -0.16 J$
where $|{\tilde{\phi}} \rangle$ is the 2D AF, i.e. spin-only part of ground state $|\phi \rangle$.
The spectrum consists of a single quasiparticle peak with a 
relatively large dispersion, cf. Fig. \ref{fig:1}(a) for the spectrum 
along the high symmetry direction $(0,0) \rightarrow (\pi, \pi)$ in the Brillouin zone.  

However, it was shown in detail in Ref. \cite{Woh11} that the above mean-field decoupling of 
spins and orbitals does not lead to a reliable description of orbital dynamics in 
a coupled spin-orbital system. Therefore, following Ref. \cite{Woh11} and using
{\it inter alia} the Jordan-Wigner transformation \cite{Jor28} for spin and orbital
degrees of freedom,\footnote{See also supplementary material in Ref. \cite{Woh11} which
discusses the validity of this mapping in the 2D case.} we map the above problem of a single orbiton in the AF and FO ground state Eq. (\ref{eq:spectral})
onto a a problem of a single hole in the AF ground state
\begin{equation} \label{eq:spectraltJ}
O({\bf k}, \omega)\!=\!\frac{1}{\pi} \lim_{\eta \rightarrow 0} 
\Im \left\langle {\tilde{\phi}}  \left|{p}^{\dag}_{{\bf k} \uparrow}   \frac{1}{\omega \!+\!
E_{{\tilde{\phi}} }  \!-\! \tilde{H} \!-\! E_z \!-\! i \eta } {p}_{{\bf k}
\uparrow}  \right|{\tilde{\phi}} \right\rangle, 
\end{equation}
described by an effective $t$--$J$ Hamiltonian
\begin{equation} \label{eq:tJ}
\tilde{H}\! = \! -\! t \!\sum_{\langle {\bf i}, {\bf j} \rangle, \sigma} ({p}^\dag_{{\bf i} \sigma}
{p}_{{\bf j} \sigma}\! +\! h.c.) 
\! +\! J \sum_{\langle {\bf i}, {\bf j} \rangle } \left({\bf S}_{{\bf i} } \cdot {\bf S}_{{\bf j}} \! +\! \frac14 {n}_{\bf i}
{n}_{\bf j}\right).
\end{equation}
Here the electron operators ${p}_{{\bf i} \sigma}$ (${p}^\dag_{{\bf i} \sigma}$) act in the restricted Hilbert space 
without double occupancies and create (annihilate) holes in the 2D AF
ground state $|{\tilde{\phi}}\rangle$. Besides, ${n}_{\bf i } =\sum_{\sigma}
n_{{\bf i}p\sigma}$ and the effective hopping parameter $t$  describing the
motion of orbital excitation is given by $t \equiv J/2$. 

The spectrum Eq. (\ref{eq:spectraltJ}) was calculated using the
self-consistent Born approximation on $32\times 32$ sites 
(which gives results in very good agreement with the exact
diagonalization studies, cf. Ref. \cite{Mar91}) and consists  
of a large quasiparticle peak and a very small incoherent spectrum, cf. Fig. \ref{fig:1}(b).
The bandwidth $W$ of the coherent orbiton peak, 
as obtained from Fig. \ref{fig:1}(b) or from perturbative calculations
cf. Ref. \cite{Mar91}, is $W \simeq 2t^2/J \simeq J /2 $
(for $t= J /2$) and thus somewhat smaller than the mean-field  one for which
$W \simeq 0.64 J$.  In more striking contrast to the mean-field
results, the dispersion has a minimum at $(\pi/ 2, \pi /2)$ instead of
$(0,0)$, as a result of the coupling between orbitons and the AF
`background'.

In many realistic cases, the spin-orbital model (\ref{eq:h}) has to be modified
to account for the different hoppings for electrons in the two active orbitals.
We have verified that the effective hopping $t$ is then no longer $t=J/2$
but instead $t= 2 t_t t_2 / U $ where $t_1$ and $t_2$ denote the bare hopping
of an electron in orbital 1 and 2 and $U$ is the on-site Hubbard repulsion. 
Furthermore, quite often the hopping of the occupied orbital 
(defined below as orbital 1) is larger, as the gain in hybridization energy contributes to its energy gain with
respect to the empty orbital, i.e., $t_1>t_2$. For example for copper 
oxides with an occupied $x^2-y^2$ orbital and empty $t_{2g}$ orbital,
$t_2 \approx t_1/2$ and thus $t=J/4$ since $J=4t^2_1/ U$. 
In the orbital-wave picture, the dispersion width is then
likewise reduced by a factor of two to $W=0.32J$, but the effect is stronger in the
interacting $t$-$J$ model, where we find $W \simeq 0.125 J$, cf. Fig. \ref{fig:1}(c)-(d).

\begin{figure}[t!]
\includegraphics[width=20pc]{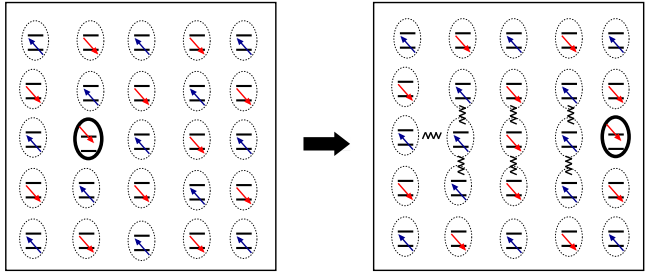}\hspace{2pc}%
\begin{minipage}[b]{16pc}\caption{\label{fig:2}Artist's view of the defects (springs) in the spin background (up and down arrows)
created by the moving orbiton (bold oval). Only if these
defects are healed by quantum fluctuations the orbiton can move
coherently -- otherwise, the orbiton is trapped in a string-like potential, see text.}
\end{minipage}
\end{figure}

\section{Discussion}
Let us now try to understand the physical origin of these 
large differences between the orbiton motion obtained 
in the mean-field approximation and in a far more exact way (i.e. via
mapping onto the effective $t$--$J$ model), cf. Fig. \ref{fig:2}. As in the mean-field case the orbiton
moves freely and does not excite any spin excitations, the whole effect
of the spin background is taken into account by renormalizing the superexchange
constant $J$ into its reduced value $J_{\rm OW}$. This stays in contrast with
the result obtained from the mapping onto the $t$--$J$ model: now
the orbiton is effectively a `hole' in the AF spin background which,
when moving, flips the spins of the 2D AF, cf. Fig. \ref{fig:2}.
In fact, the further the orbiton moves the more spin flips (defects
in 2D AF) it excites which leads to the so-called string-like
potential acting on the orbiton, cf. Fig. \ref{fig:2}. The orbiton can
then move coherently 
only due to the presence of quantum spin fluctuations in the ground state which 
`heal' these defects, `attach' to the orbiton and jointly move
as an orbiton quasiparticle on a renormalized scale 
(cf. Ref. \cite{Sch88, Mar91} with the hole in 2D AF 
moving effectively as a spin polaron).
We also note that this result is entirely different from that in the
1D case~\cite{Woh11}, where the orbital 
flip excitation fractionalizes into free 'spinon' and 'orbiton'
quasiparticles, similar to decay of a doped hole into spinon and holon.

\section{Conclusions}
In summary, we showed that a single orbital excitation in the 2D AF and FO ground state
moves similarly to a hole in the 2D AF state. Therefore, it can only be mobile
by coupling to the spin fluctuations which leads to a minimum in the orbiton
dispersion for the $(\pi/2 , \pi/2)$ point in the Brillouin zone 
and to the very strong renormalization of the orbiton bandwidth. In particular, 
in the realistic case which can correspond to the $dd$ excitations in copper oxides, the orbiton
bandwidth is of the order of $\sim 0.1 J$ which explains why typically such
$dd$ excitations in the AF ground state are nearly dispersionless and can 
be well modelled using quantum chemistry calculations on small clusters \cite{Hoz11}.

\ack
We acknowledge support of the Alexander von Humboldt Foundation (K.W.) 
and the DFG Emmy Noether Program (M.D.).


\section*{References}
\begin{thebibliography}{9}
\bibitem{Col01} Coldea R {\it et al.} 2001 {\it Phys. Rev. Lett.} {\bf 86} 5377
\nonum Walters A C {\it et al.} 2009 {\it Nat. Phys.} {\bf 5} 867

\bibitem{Gru02} Gr\"uninger M {\it et al.} 2002 {\it Nature} \textbf{418} 39
\nonum Ulrich C {\it et~al.} 2009 {\it Phys. Rev. Lett.} {\bf 103} 107205

\bibitem{vdB98} van den Brink J, Stekelenburg W, Khomskii D I, Sawatzky G A and Kugel K I 1998
                {\it Phys. Rev.} B {\bf 58} 10276
\nonum   van den Brink J, Horsch P, Mack F and Ole\'s A M 1999
                {\it Phys. Rev.} B {\bf 59} 6795

\bibitem{Woh11}
Wohlfeld K, Daghofer M, Nishimoto S, Khaliullin G and van den Brink J 2011 
{\it Phys.~Rev.~Lett.} {\bf 107} 147201

\bibitem{KK1982}
Kugel K I and Khomskii D I 1982 {\it Sov. Phys. Usp.} {\bf 25} 231

\bibitem{Kha97} 
Khaliullin G and Oudovenko V 1997 {\it Phys.~Rev.}~B {\bf 56} 14243

\bibitem{Ole05} 
Ole\'s A M, Khaliullin G, Horsch P and Feiner L F 
2005 {\it Phys.~Rev.}~B \textbf{72} 214431

\bibitem{Woh09}
Wohlfeld K, Ole\'s AM and Horsch P 2009 {\it Phys.~Rev.}~B 
{\bf 79} 224433

\bibitem{Jor28}
Jordan P and Wigner E 1928 {\it Z.~Physik} {\bf 47} 631

\bibitem{Mar91} 
Martinez G and Horsch P 1991 {\it Phys.~Rev.}~B {\bf 44} 317

\bibitem{Sch88} 
Schmitt-Rink S, Varma C M and Ruckenstein A E 1988 {\it Phys. Rev. Lett.} {\bf 60} 2793

\bibitem{Hoz11}
Hozoi L, Siurakshina L, Fulde P and van den Brink J 2011 
{\it Sci. Rep.} {\bf 1} 65



\end{thebibliography}
\end{document}